\documentclass{article}
\makeatletter
\@addtoreset{equation}{section}

\makeatother

\begin{document}

\begin{flushright}
Mar 2005

KIAS-P05017
\end{flushright}

\begin{center}

\vspace{5cm}

{\Large Tachyons in Compact Spaces}

\vspace{2cm}

Takao Suyama \footnote{e-mail address : suyama@kias.re.kr}

\vspace{1cm}

{\it School of Physics, Korea Institute for Advanced Study,}

{\it 207-43, Cheongnyangni 2-dong, Dongdaemun-gu, Seoul 130-722, Korea}

\vspace{4cm}

{\bf Abstract} 

\end{center}

We discuss condensations of closed string tachyons localized in compact spaces.
Time evolution of an on-shell condensation is naturally related to the worldsheet RG flow.
Some explicit tachyonic compactifications of Type II string theory is considered, and some of
them are shown to decay into supersymmetric theories known as the little string theories.
\newpage

\vspace{1cm}

\section{Introduction}

\vspace{5mm}

Condensations of closed string tachyons has been studied recently. 
There are review papers \cite{review1}\cite{review2} on this subject. 
Up to now, most of investigations have been done for 
the situations in which the relevant part of the target spacetime 
is non-compact, and tachyons whose condensations are investigated are localized in that part of the 
spacetime. 
Typical examples for such situations are string theories on the Melvin, or twisted circle background 
\cite{CostaGutperle}, and on non-compact orbifolds \cite{APS}. 

There are advantages to consider such tachyons. 
Localized tachyons are, by definition, localized in a small region of spacetime. 
For a Melvin case, which is described as $({\bf C}\times S^1)/{\bf Z}_N$, 
tachyons come from strings winding on the $S^1$ direction, and they are localized 
at the origin of the ${\bf C}$ in this geometry. 
For the orbifold case, for example ${\bf C}/{\bf Z}_N$, 
tachyons come from twisted sectors which represent strings winding around the 
fixed point. 
In these cases, it is intuitively clear that the presence of the tachyons indicates some kind of 
instability 
of a small part of the geometry to which the tachyons are associated. 
In fact, it is argued that the twisting of $S^1$ disappears for the Melvin case, and the fixed point 
is smoothed out for the orbifold case, via a tachyon condensation. 
In the above examples, these phenomena are described as reductions of $N$ to smaller values. 
There is also a technical advantage to analyze the tachyon condensations in these situations. 
Usually, a tachyon condensation is assumed to be 
described by a worldsheet RG flow induced by the corresponding 
tachyon vertex operator. 
Since the operator is relevant, the central charge of the theory decreases in general \cite{c-th}. 
In the above-mentioned cases, the central charge of the theory does not decrease along the RG flow, 
and therefore, the modification of the theory due to the tachyon condensation is a mild one. 

If one consider a condensation of a bulk tachyon \cite{bulk1}\cite{bulk2}\cite{bulk3}, 
or a localized tachyon living in a compact 
space, 
it seems that the central charge of the theory inevitably decreases. 
For example, some part of spacetime might completely disappear, and the dimensionality of the 
spacetime might decrease. 
If such things happen, it would be difficult to interpret the resulting theory as a compactification 
of a string theory in the critical dimensions, and therefore, the spacetime dynamics of the tachyon 
condensation would be unclear
\footnote{Recently, there appeared a paper \cite{recent} 
which claimed that a closed string tachyon 
condensation not always decreases the dimensionality of spacetime, although it might change the 
topology of the spacetime.}. 

In this paper, we would like to discuss localized closed string tachyons living in compact spaces. 
What will be discussed below are {\it on-shell} time evolutions of tachyon condensations. 
These time evolutions can be naturally related to the worldsheet RG flows by identifying the time 
direction with the Liouville direction, which is a result of the well-known relation between 
critical string theories and non-critical string theories. 
Then the late-time evolution of a system can be obtained by investigating the IR fixed point of 
the RG flow of the non-critical theory. 
The obtained IR theory can be again regarded as a critical string theory, and this theory would 
describe the 
end-point of the tachyon condensation. 
In this way, a kind of tachyonic compactifications of Type II string theory is considered, and some 
of them are shown to have a non-trivial supersymmetric, i.e. stable, end-point of the decay. 

This paper is organized as follows. 
In section \ref{time&RG}, 
the relation between the time evolution and the worldsheet RG flow is explained. 
A rolling tachyon background \cite{rolling1}\cite{rolling2} is also understood in this way. 
In section \ref{examples}, 
some of explicit tachyonic compactifications of Type II string theory are discussed. 
It is also shown that Type II string theory on a three-dimensional compact space decays into the 
little string theory \cite{LST0}\cite{LST}, which means that this decay has a stable end-point. 
Section \ref{discussion} is devoted to discussion.

\vspace{1cm}

\section{Tachyon dynamics and worldsheet RG} \label{time&RG}

\vspace{5mm}

\subsection{Liouville direction as time}

\vspace{5mm}

Let us recall a well-known 
relation between a critical string theory and a non-critical string 
theory, i.e. 
a CFT coupled to the 
two-dimensional (super)gravity. 
For simplicity, we will first review such a relation for bosonic case. 
For a review, see e.g. \cite{GinspargMoore}. 

Consider a CFT with central charge $c_m$ coupled to the two-dimensional gravity. 
This theory is described by the following partition function
\begin{equation}
Z = \int {\cal D}g{\cal D}X\ e^{-S_m(X;g)}, 
   \label{Z}
\end{equation}
where $g=g_{\alpha\beta}$ is the metric on the worldsheet, and $X$ collectively represents matter 
fields of the CFT. 
As is well-known, any metric $g$ can be transformed by a diffeomorphism $f$ to 
\begin{equation}
f^*g = e^{\varphi}\hat{g}(\tau), 
\end{equation}
where $\hat{g}(\tau)$ is a representative metric of a conformal 
equivalence class of metrics specified by the 
moduli parameters $\tau=\{\tau^i\}$. 
This fact implies that 
the path-integral on $g$ can be divided into three integrals; the path-integral on $\varphi$, 
the integrals on the moduli parameters $\tau$ and an 
integral along metrics which are related to $f^*g$ by a diffeomorphism, accompanied by the FP 
determinant corresponding to this gauge choice. 
Since the action $S_m(X;g)$ is invariant under diffeomorphisms, 
by definition, the third integral provides 
the volume of the group of diffeomorphisms which is an irrelevant constant. 
After rewriting the path-integral and dividing out the gauge volume, the partition function (\ref{Z}) 
is written as 
\begin{equation}
Z = \int d\tau{\cal D}\varphi{\cal D}b{\cal D}c{\cal D}X\ 
    e^{-S_m(X;\hat{g})-S_{gh}(b,c;\hat{g})}, 
\end{equation}
where $S_{gh}(b,c;g)$ is the action of the FP ghosts $b,c$. 
It has been used that both $S_m(X,g)$ and $S_{gh}(b,c;g)$ are Weyl-invariant. 

Note that the path-integral measures are defined in terms of $e^\varphi\hat{g}(\tau)$, that is, there 
are implicit dependences on $\varphi$. 
It is usually assumed \cite{D}\cite{DK} that the dependence on $\varphi$ of the measures is 
\begin{equation}
{\cal D}\varphi{\cal D}b{\cal D}c{\cal D}X 
= e^{-S_L(\varphi;\hat{g}(\tau))}\hat{{\cal D}}\varphi\hat{{\cal D}}b\hat{{\cal D}}c\hat{{\cal D}}X, 
\end{equation}
where $\hat{{\cal D}}\varphi$ etc. are defined in terms of $\hat{g}(\tau)$. 
The Liouville action $S_L(\varphi;\hat{g}(\tau))$ is 
\begin{equation}
S_L(\varphi;\hat{g}(\tau)) 
= \frac1{8\pi}\int d^2\sigma\sqrt{\hat{g}(\tau)}\Bigl[\ \frac{25-c_m}{12}(\hat{\nabla}\varphi)^2
  +\frac{25-c_m}6\hat{R}\varphi\ \Bigr], 
\end{equation}
which are determined by requiring that, as it should be, the final form of the partition function 
\begin{equation}
Z = \int d\tau\hat{{\cal D}}\varphi\hat{{\cal D}}b\hat{{\cal D}}c\hat{{\cal D}}X\ 
    e^{-S_m(X;\hat{g})-S_{gh}(b,c;\hat{g})-S_L(\varphi;\hat{g})}
\end{equation}
is invariant under 
\begin{equation}
\hat{g} \to e^{\omega}\hat{g}, \hspace{5mm} \varphi \to \varphi-\omega, 
\end{equation}
where $\omega$ is an arbitrary function on the worldsheet. 
It is better to rescale $\varphi$ so that its kinetic term has an appropriate form, 
for example, 
\begin{equation}
S_L(\varphi;\hat{g}(\tau)) 
= \frac1{8\pi}\int d^2\sigma\sqrt{\hat{g}(\tau)}\Bigl[\ (\hat{\nabla}\varphi)^2
  +Q\hat{R}\varphi\ \Bigr], 
\end{equation}
where 
\begin{equation}
Q = \sqrt{\frac{25-c_m}3}. 
\end{equation}
Note that the resulting theory is conformal and its central charge is 
\begin{equation}
c = c_m+1+3Q^2-26 = 0. 
\end{equation}
Therefore, 
the above argument shows that a gauge-fixed action of the CFT coupled to the two-dimensional gravity 
can be regarded as the action of a critical string theory with a linear dilaton background. 

In the case with $c_m>25$, $Q$ is purely imaginary. 
This is simply because $\varphi$ has been rescaled by an imaginary quantity. 
If an appropriate rescaling is done, one would instead obtain 
\begin{equation}
S_L(\varphi;\hat{g}(\tau)) 
= \frac1{8\pi}\int d^2\sigma\sqrt{\hat{g}(\tau)}\Bigl[\ -(\hat{\nabla}\varphi)^2
  -Q'\hat{R}\varphi\ \Bigr], 
\end{equation}
where 
\begin{equation}
Q' = \sqrt{\frac{c_m-25}3}. 
\end{equation}
Note that the total central charge is again $c=0$. 
In this case, the kinetic term of $\varphi$ has the wrong sign, indicating that the Liouville 
direction is regarded as the {\it time direction}. 
The interpretation of the Liouville field as the time direction was employed in e.g. 
\cite{time1}\cite{time2}\cite{time3}. 

In a limiting case $c_m=25$, in which the rescaling would need some regularization, the dilaton 
gradient $Q$ vanishes, and the resulting theory is the usual critical string theory compactified on 
a 26-dimensional Lorentzian manifold 
${\bf R}\times M^{25}$ where ${\bf R}$ is the Liouville (time) direction and 
$M^{25}$ is a 25-dimensional Euclidean manifold corresponding to the CFT, provided that the CFT is 
unitary. 
If the $c_m=25$ case is regarded as a limiting case of $c_m<25$, there might seem to be no problem on 
regarding $\varphi$ as spatial direction. 
However, 
it would be reasonable to require that 
an operator $A=\int d^2\sigma\sqrt{\hat{g}(\tau)}e^{\gamma\varphi}$ in the gauge-fixed action, 
which should correspond to the area of the worldsheet $\int d^2\sigma\sqrt{g}$, 
should be a real quantity. 
Since $e^{\gamma\varphi}$ should have weight $(1,1)$ so that $A$ is well-defined, 
$\gamma$ should satisfy 
\begin{equation}
-\frac12\gamma^2 = 1.  
\end{equation}
Therefore, $A$ is real if $\varphi$ is Wick-rotated, and this rotation 
changes the sign of the kinetic 
term of $\varphi$. 

\vspace{5mm}

It is straightforward to generalize this kind of relation to the case of a 
SCFT coupled to the two-dimensional supergravity \cite{super}. 
The resulting gauge-fixed theory is the SCFT plus ${\cal N}=1$ super-Liouville theory. 
To regard this gauge-fixed theory as Type II string theory or Type 0 one, 
one has to consider carefully about the GSO-projection. 
The fermion for the time direction comes from the two-dimensional gravitino, and it might 
look quite different 
from fermions in SCFT. 
Therefore, naively it seems that 
the GSO-projection in the SCFT has nothing to do 
with the gravitino. 
However, since the original theory is a supergravity, there is a coupling of the gravitino to the 
supercurrent of the SCFT. 
Due to this coupling, the $(-1)^{F_{L(R)}}$ of the SCFT 
are symmetries of the total theory if $(-1)^{F_{L(R)}}$ also 
flip 
the sign of the gravitino. 
Therefore, if one starts from the Type II (0) SCFT coupled to the two-dimensional supergravity, 
then its gauge-fixed theory is the Type II (0) string theory in a linear dilaton background, 
respectively.

\vspace{5mm}

\subsection{Relevant perturbations}

\vspace{5mm}

Consider again a bosonic CFT coupled to the two-dimensional gravity. 
Suppose that the CFT has a relevant operator $V$ with weight $\Delta<1$. 
One can consider a deformation of the original CFT by the operator $V$ as follows, 
\begin{equation}
S'_m(X;g) = S_m(X;g) + \lambda\int d^2\sigma\sqrt{g}\ V. 
   \label{deform}
\end{equation}
The deformed matter theory is not a CFT, and 
the operator $V$ induces an RG flow. 
We assume that there is an IR fixed point of the RG flow which is described by the action 
$S^*_m(X;g)$. 
It is expected that, even for theories coupled to two-dimensional gravity, there would be 
a modified version 
of the c-theorem, and the degrees of freedom of the theories would decrease along the RG flow 
\cite{RG}. 
In the following, we will argue that the IR fixed point describes the end-point of the 
condensation of the tachyon $V$. 

\vspace{3mm}

In a similar way shown in the previous subsection, the deformed theory (\ref{deform}) can be 
related to a critical string theory via the gauge fixing of diffeomorphisms. 
The gauge-fixed action is 
\begin{equation}
S' = S_m(X;\hat{g}(\tau)) + S_{gh}(b,c;\hat{g}(\tau)) + S_L(\varphi;\hat{g}(\tau)) 
    +\lambda\int d^2\sigma\sqrt{\hat{g}(\tau)}e^{\alpha\varphi}V, 
\end{equation}
where $\alpha$ satisfies 
\begin{equation}
\Delta-\frac12\alpha(\alpha-Q) = 1. 
   \label{exponent}
\end{equation}
For the case $c_m=25$, one obtains $\alpha=\pm i\sqrt{2(1-\Delta)}$, and therefore 
\begin{equation}
S' = S_m(X;\hat{g}(\tau)) + S_{gh}(b,c;\hat{g}(\tau)) + S_L(\varphi;\hat{g}(\tau)) 
    +\lambda\int d^2\sigma\sqrt{\hat{g}(\tau)}e^{\sqrt{2(1-\Delta)}\varphi}V, 
       \label{timedependent}
\end{equation}
after the Wick rotation of $\varphi$. 
The sign in $\alpha$ is included in $\varphi$. 

The fact that the action (\ref{timedependent}) is conformally invariant means that this action 
describes a critical string theory living in an on-shell time dependent background, since 
$\varphi$ is regarded as describing the time direction. 
The perturbation term becomes large as $\varphi\to+\infty$. 
That is, this background describes an {\it on-shell tachyon condensation} by identifying the increase 
of $\varphi$ as the increase of time. 
Therefore, by taking $\varphi$ very large, one can find the end-point of the tachyon 
condensation, in principle. 

If the operator $e^{\sqrt{2(1-\Delta)}\varphi}V$ is exactly marginal, the late-time behavior of the 
tachyon condensation is 
described by the same 
theory (\ref{timedependent}) with a large $\varphi$. 
In general, this operator is not marginal for large $\lambda$, and therefore for large $\varphi$. 
In fact, the exponent $\alpha$ is chosen so that 
the operator $e^{\alpha\varphi}V$ is marginal for $\lambda=0$, but this does not 
necessarily imply that it is marginal for a general $\lambda$. 
If the perturbation is not exactly marginal, the only thing one can say is that 
the action (\ref{timedependent}) approximately describes 
a string theory in the on-shell time dependent tachyon background 
when $e^{\sqrt{2(1-\Delta)}\varphi}$ is small, or in other words, at the beginning of the 
condensation. 
However, it would not be a good description for the final stage of the condensation. 
In the large $\varphi$ region in which 
the perturbation 
becomes large, the backreaction of the condensation to the background 
geometry would not be negligible. 

Recall that 
the Liouville field $\varphi$ originally comes from the conformal factor of the worldsheet metric $g$. 
Due to this fact, a shift of $\varphi$, in other words time evolution, 
should correspond to a change of the length scale of the worldsheet 
in the original two-dimensional gravity. 
This provides a natural relation between the dynamics of the tachyon condensation and the 
RG flow of the two-dimensional gravity \cite{RG}. 
To set the scale of the latter theory, we fix the area of the worldsheet $A$, which corresponds in 
the gauge-fixed theory to inserting the delta-function 
\begin{equation}
\delta\Bigl( A-\int d^2\sigma\sqrt{\hat{g}(\tau)}e^{\sqrt{2}\varphi} \Bigr). 
\end{equation}
It would be obvious from the definition that the small $A$ limit corresponds to the UV limit of the 
RG flow, and the large $A$ limit corresponds to the IR limit. 
Since it is shown above that, around the UV limit, 
the two-dimensional gravity (\ref{deform}) 
describes the early stage of the tachyon condensation, it would be 
reasonable to assume that the same theory in the IR limit would describe the final stage 
of the condensation. 

\vspace{3mm}

Suppose that the IR fixed point described by $S^*_m(X;g)$ has the central charge $c_{IR}$. 
Then, the gauge-fixed action for the IR fixed point is 
\begin{equation}
S^* = S^*_m(X;\hat{g}(\tau)) + S_{gh}(b,c;\hat{g}(\tau)) + S_L(\varphi;\hat{g}(\tau)), 
   \label{IR}
\end{equation}
where the dilaton gradient $Q_{IR}$ is now 
\begin{equation}
Q_{IR} = \sqrt{\frac{25-c_{IR}}{3}}. 
\end{equation}
In general, $c_{IR}$ would be smaller than $c_m$. 
If the UV theory $S_m(X;g)$ has $c_m=25$, one obtains $Q_{IR}>0$. 
Note that in the IR theory, the Wick-rotation of $\varphi$ will change the total central charge, since 
the Wick-rotation induces $Q_{IR}\to iQ_{IR}$. 
Therefore, the Liouville field $\varphi$ cannot be Wick-rotated without changing the theory itself, 
and the time direction has to be chosen from other directions. 
This issue of the choice of the time direction is briefly discussed in section \ref{discussion}. 

\vspace{3mm}

Let us summarize the above arguments, 
and provide a prescription to investigate a tachyon condensation. 
Suppose that critical 
strings are living in a background which provides a tachyon in the mass spectrum. 
There would exist a CFT coupled to the two-dimensional gravity (\ref{Z}) whose gauge-fixed theory is 
equivalent to the tachyonic critical string theory. 
To consider a condensation of the tachyon whose corresponding operator in CFT is $V$, one 
should consider a deformed theory (\ref{deform}), and 
investigate the IR limit of the RG flow induced by $V$. 
The critical 
string theory describing the late-time behavior of the on-shell tachyon condensation is described 
by the gauge-fixed action of the IR fixed point, i.e. (\ref{IR}). 
The resulting string theory has a linear dilaton background in general. 

\vspace{3mm}

This prescription can apply to the problem on open string tachyon condensations in bosonic string 
theory \cite{Sen}. 
Consider a homogeneous condensation of the open string tachyon. 
The matter part of the two-dimensional gravity theory one has to consider is the CFT of 25 free bosons. 
Since the tachyon is the ground state of the matter theory, the operator to be considered should be 
the unit operator ${\bf 1}$. 
The deformation term is $\int_{boundary}d\xi\ {\bf 1}$. 
The corresponding term in the gauge-fixed action is 
\begin{equation}
\int_{boundary}d\xi\ e^{\varphi/\sqrt{2}}. 
   \label{halfS}
\end{equation}
As a result, one obtains the theory of half-S-brane \cite{rolling1}\cite{rolling2}. 
The worldsheet RG flow coincides with the 
shift of $\varphi$.
Since the deformation (\ref{halfS}) 
is the boundary one, the matter central charge does not 
change along the RG flow, and no dilaton gradient is produced by the tachyon condensation. 
The prescription also applies to the tachyon condensation in which the tachyon has a non-zero 
momentum in spatial directions.

\vspace{1cm}

\section{Condensations of tachyons in compact spaces} \label{examples}

\vspace{5mm}

In this section, we apply the prescription for investigating tachyon condensations proposed in the 
previous section to some explicit examples of string theories. 
In particular, there is a Type II string theory with localized tachyons in a compact space, and the 
condensation of the tachyons would lead it to another supersymmetric theory known as the little 
string theory. 

\vspace{5mm}

\subsection{Type II string theory on $T^2/{\bf Z}_3$}

\vspace{5mm}

The first example is Type II string theory compactified on $T^2/{\bf Z}_3$. 
The torus is defined by the identifications 
\begin{equation}
X \sim X+2\pi mR+2\pi nR\zeta, 
\end{equation}
where $X=X^8+iX^9,\ m,n\in{\bf Z}$ and $\zeta=e^{2\pi i/3}$. 
A generator $g$ of ${\bf Z}_3$ acts on $X$ as 
\begin{equation}
g\cdot X = \zeta X. 
\end{equation}
As discussed in \cite{APS}, 
there are two such orbifold theories depending on the action of ${\bf Z}_3$ on 
spacetime fermions. 
In the following, a generator $g$ of ${\bf Z}_3$ is chosen such that 
\begin{equation}
g = \exp(2\pi iJ_{89}/3)\exp(2\pi iJ_{89}), 
\end{equation}
where $J_{89}$ is the generator of rotations in 89-plane. 
The definition of ${\bf Z}_3$ is so chosen not to obtain a bulk tachyon in the mass 
spectrum. 
This two-dimensional orbifold has three fixed points, 
\begin{equation}
X_{fix} = 0,\ \frac{2\pi R}3(2+\zeta),\ \frac{2\pi R}3(1+2\zeta), 
\end{equation}
each of which has two localized tachyons with 
\begin{equation}
\frac{\alpha'}4M^2 = -\frac13,
\end{equation}
in the twisted sectors. 
This orbifold theory has ${\cal N}=2$ worldsheet supersymmetry, and a 
half of the tachyons are chiral primary 
states, while 
the other half are anti-chiral primaries. 

It is known \cite{VW} that this particular orbifold can be described by an ${\cal N}=2$ 
Landau-Ginzburg model which is 
defined by the superpotential 
\begin{equation}
W(X,Y,Z) = X^3+Y^3+Z^3, 
   \label{LG}
\end{equation}
when the 
complexified K\"ahler structure of $T^2$ has an appropriate value. 
The fact that the LG model has three superfield $X,Y,Z$ corresponds to the presence of three fixed 
points in the orbifold. 
The superfield $X$ corresponds to a tachyon in a twisted sector of the fixed point related to $X$. 
The correspondence of low-lying states between the orbifold and the LG model is shown in appendix 
\ref{LGspectrum}. 

The correspondence can be also seen by the gauged linear sigma model \cite{GLSM}. 
The model relevant here 
is a two-dimensional ${\cal N}=(2,2)$ $U(1)$ gauge theory coupled to four superfields $X,Y,Z,P$ 
whose $U(1)$ charges are $+1,+1,+1,-3$, respectively. 
This gauge theory has the superpotential 
\begin{equation}
W(X,Y,Z,P) = P(X^3+Y^3+Z^3), 
\end{equation}
and a (complexified) FI term. 
In the Calabi-Yau phase, in which the FI parameter takes a large positive value, the dynamics of 
massless degrees of freedom is described by a non-linear sigma model whose target space is the 
curve defined by 
\begin{equation}
X^3+Y^3+Z^3=0
   \label{curve}
\end{equation}
in $CP^2$. 
This curve is equivalent to the curve 
\begin{equation}
y^2z=\frac43x^3-\frac13z^3
   \label{curve2}
\end{equation}
which is related to (\ref{curve}) by $x=X,\ y=Y-Z,\ z=-Y-Z$. 
This equation shows that the modulus of this torus is $\tau=e^{2\pi i/3}$. 
On the other hand, 
in the LG phase, in which the FI parameter takes a large negative value, $P$ has a non-zero 
expectation value, and the resulting theory is described by the LG model with the superpotential 
$W(X,Y,Z,\langle P\rangle)$. 
However, there is a residual ${\bf Z}_3$ gauge symmetry, and the LG model should be modded out by 
this symmetry. 
In this way, the gauged linear sigma model provides a relation between the torus $T^2$ with modulus 
$\tau=e^{2\pi i/3}$ and the LG orbifold. 
It is mentioned in \cite{Vafa} that 
taking a ${\bf Z}_3$ orbifold of the $T^2$ is equivalent to eliminating the effect 
of orbifolding in the LG orbifold. 

\vspace{3mm}

Now it is easy to find the end-point of a tachyon condensation in $T^2/{\bf Z}_3$ since the RG flow 
of the LG model is known \cite{VW}. 
First, consider a condensation of a tachyon corresponding to the superfield $Z$. 
The deformed theory has the LG superpotential 
\begin{equation}
W'(X,Y,Z) = X^3+Y^3+Z^3+\lambda_ZZ. 
\end{equation}
In the IR limit, the term $Z^3$ becomes irrelevant and the linear term $Z$ remains. 
This means that the degrees of freedom of $Z$ decouple from the theory. 
The superpotential of the IR theory is therefore 
\begin{equation}
W_{IR}(X,Y) = X^3+Y^3. 
   \label{LG'}
\end{equation}
Since this LG model has the central charge $c_{LG}=2<c_{UV}$, 
the IR background has a linear dilaton. 
This IR theory still has relevant operators $X,\ Y$. 
Therefore, this theory may possibly decay further 
due to condensations of these tachyons, and finally the LG 
part of the theory would become trivial. 
Then the target space of the final theory would be a six-dimensional one. 

It is not clear whether this final theory, or even the intermediate theory like (\ref{LG'}), has an 
interpretation as a string theory in ten dimensions. 
It might be possible to regard such a theory as a string theory living in a non-compact space in which 
the Liouville direction would describe a non-compact direction. 
In the next example, it is shown that the end-point of a tachyon condensation is a theory related to 
NS5-branes, i.e. the little string theory.

\vspace{5mm}

\subsection{Type II string theory on $(T^2/{\bf Z}_3\times S^1)/{\bf Z}_3$}

\vspace{5mm}

It is obvious that the LG model (\ref{LG}) has a symmetry 
\begin{equation}
g_X\ :\ X \to \zeta X, \hspace{5mm} Y,Z \mbox{ fixed},
\end{equation}
and the similar symmetries $g_Y,\ g_Z$ for $Y$ and $Z$, respectively. 
The action of $g_X$ on the geometry (\ref{curve}) or (\ref{curve2}) is also obvious. 
One can consider a ${\bf Z}_3$ orbifold of $T^2/{\bf Z}_3\times S^1$, in which a generator $g'$ of 
the ${\bf Z}_3$ acts on the LG model fields as $g_X$ and 
\begin{equation}
g'\ :\ X^7\ \to\ X^7+\frac{2\pi r}3, 
\end{equation}
where $X^7$ is the bosonic coordinate for the $S^1$, and $r$ is the radius of the $S^1$. 
In this orbifolded theory, the tachyon $X$ is projected out. 

One can consider a condensation of the tachyons $Y,\ Z$ in the same way as in the previous subsection. 
The IR fixed point of the flow describes a critical string theory living in 
\begin{equation}
{\bf R}^6\times {\bf R}\times(S^1\times (\mbox{$k=1$ minimal model}))/{\bf Z}_3, 
\end{equation}
which is, for a suitable choice of $r$, 
the string theory propagating near NS5-branes \cite{CGS}, or in other words, 
the holographic dual CFT \cite{LST} of the 
worldvolume theory of NS5-branes, i.e. the little string theory \cite{LST0}. 
This theory has 16 spacetime supercharges and definitely stable. 
Therefore, this decay of Type II string theory on $(T^2/{\bf Z}_3\times S^1)/{\bf Z}_3$ is an 
explicit example of condensations of localized closed string tachyons in compact spaces 
which have a stable and 
non-trivial end-point. 

It might seem to conflict with the c-theorem \cite{c-th} that 
the target space of the IR fixed point theory is ten-dimensional, since the ${\bf Z}_3$ 
orbifold part is equivalent to $S^3$. 
However, in the spacetime sense, the number of degrees of freedom decreases after the tachyon 
condensation, since the IR theory is 
supposed to be equivalent to the worldvolume theory of NS5-branes, the bulk degrees of freedom being 
decoupled. 

It will be interesting to point out that the UV geometry $(T^2/{\bf Z}_3\times S^1)/{\bf Z}_3$ can be 
regarded as a compact analog of the Melvin geometry, or the twisted circle \cite{CostaGutperle}. 
It can be also seen as a duality twist \cite{dualtwist}. 
In the non-compact case, it is argued that a string theory in this kind of space would decay into the 
supersymmetric flat background due to a condensation of a closed string tachyon. 
In the compact case discussed here, the IR background is also supersymmetric, although it 
is not flat.

\vspace{5mm}

\subsection{Other tachyonic backgrounds}

\vspace{5mm}

There are two more $T^2$ orbifolds which can be described by LG models. 
\begin{equation}
\begin{array}{ccc}
T^2/{\bf Z}_4 & : & W(X,Y,Z) = X^4+Y^4+Z^2, \\
T^2/{\bf Z}_6 & : & W(X,Y,Z) = X^6+Y^3+Z^2.
\end{array}
\end{equation}
As discussed in \cite{APS}, each of these orbifolds has a bulk tachyon. 
As long as the vev of the bulk tachyon is tuned to be zero, one can argue condensations of the other 
localized tachyons in a similar way as above. 

One can also 
consider a Melvin-type orbifold of the above $T^2$ orbifold times $S^1$, 
like that discussed in the previous subsection. 
Such Melvin-type orbifolds would decay again into the little string theory. 
The level of the minimal model in the IR theory, or the number of NS5-branes, can be 3, 4 and 6, 
depending on which LG superfield is used to obtain the Melvin-type orbifold. 

There is a proposal \cite{LST} 
of a holographic description of some compactified NS5-branes in terms of 
worldsheet CFT's. 
An example is Type II string theory on the target spacetime 
\begin{equation}
{\bf R}^4\times {\bf R}\times S^1\times (\mbox{LG model}), 
\end{equation}
where the LG model is defined by the superpotential 
\begin{equation}
W = Z_1^2+Z_2^2+Z_3^2+Z_4^2, 
\end{equation}
and this worldsheet theory is proposed to describe NS5-brane on the conifold. 
It is very natural to expect that there is a tachyonic compactification of a ten-dimensional string 
theory which decays into this 
little string theory via a tachyon condensation. 
If such a tachyonic theory actually exists, it is a Type II string theory whose target spacetime is 
${\bf R}^5\times S^1\times M$ where $M$ is an internal manifold described by a CFT with central 
charge $c=6$. 
If this CFT is equivalent to a LG model, then one can do the similar analysis. 
There are various LG models with $c=6$, and therefore, naively, there would be various tachyonic 
compactifications of Type II string theory which could decay into the lower-dimensional little string 
theory. 
It will be very interesting to investigating a pattern of such decay processes and relate many 
critical string theories to little string theories, which will hopefully be reported elsewhere.

\vspace{1cm}

\section{Discussion}  \label{discussion}

\vspace{5mm}

In this paper, we have discussed condensations of closed string tachyons in compact spaces. 
Based on the interpretation of the Liouville direction as the 
time direction, a time evolution of an on-shell 
tachyon condensation in a critical string theory is related to a worldsheet RG flow of the 
corresponding two-dimensional (super)gravity coupled to matter. 
In the case where the worldsheet RG flow is known, like ${\cal N}=2$ LG models, 
it is possible to find an 
end-point of the condensation which typically has a linear dilaton background. 
It is shown that there is a particular tachyonic string theory, Type II string theory on 
$(T^2/{\bf Z}_3\times S^1)/{\bf Z}_3$, which decays into a supersymmetric string theory, i.e. 
the little string theory. 

As briefly mentioned before, there is no contradiction to the c-theorem. 
The RG flow discussed in this paper is that of a two-dimensional matter theory coupled to gravity, 
and the central charge of the matter theory actually decreases since the flow is induced by a 
relevant operator. 
In terms of the critical string theory, which is a gauge fixed form of the two-dimensional gravity, 
the total central charge is unchanged since the relevant operator is dressed and becomes a 
(not always exactly) marginal operator. 
The dressed operator describes a time-dependent on-shell background, and therefore, the total central 
charge does not change during the time evolution. 
It would be helpful to point out that the RG flow discussed in this paper is different from the RG 
flow which is usually discussed in the context of tachyon condensation. 
In the latter, since 
the Liouville field is just a usual scalar field, it is renormalized multiplicatively, 
but in the former, the `renormalization' of the Liouville field is additive. 

Note that the number of spacetime degrees of freedom can, in some sense, change while the central 
charge is kept fixed. 
In fact, a naive counting of spacetime dimensions is not a definite notion, especially in string 
compactifications constructed by using LG models. 
For example, $(T^2/{\bf Z}_3\times S^1)/{\bf Z}_3$ is a three-dimensional space. 
On the other hand, the corresponding LG model $(S^1\times X^3)/{\bf Z}_3+Y^3+Z^3$ can be regarded 
as a space consists of $S^3$ ($(S^1\times X^3)/{\bf Z}_3$ part) and the remaining part which is 
possibly a discrete space, and therefore, this can be regarded as a $(3+\alpha)$-dimensional space. 
This example suggests that the change of the number of degrees of freedom is not always related to 
the change of dimensionality of spacetime. 

As pointed out, the space $(T^2/{\bf Z}_3\times S^1)/{\bf Z}_3$ resembles a Melvin background, 
and it is tempting to relate the decay of the latter background to the decay of the former 
in a decompactification limit. 
If the radius, or more generally the complexified K\"ahler structure, of the $T^2$ changes, the 
correspondence to the LG model is lost, and the analysis of RG flows would become difficult. 
However, it would be a natural guess that $(T^2/{\bf Z}_3\times S^1)/{\bf Z}_3$ for 
large $T^2$ also decays into $S^3$ with large radius. 
If this is true, the tachyon condensation process in the decompactification limit coincides with 
what is expected for Melvin backgrounds. 

The appearance of the little string theory as a result of a tachyon condensation may be natural. 
Consider the decay in terms of a spacetime effective theory. 
The end-point of the decay would be described by a non-trivial configuration of background fields, 
including tachyons, and this configuration could have a finite energy density. 
Since the decay is a phenomenon in the closed string theory, the action of the effective theory, 
and therefore the energy of the system, is 
multiplied by $1/g_{st}^2$, suggesting that the background may be related to NS5-branes. 

It is very interesting if there exists an exact CFT describing the whole process of the tachyon 
condensation. 
Such a theory would interpolate the UV theory and the IR theory which is proposed in this paper. 
If this interpolating theory has a sigma model expression, its background field configuration would be 
a very complicated one. 
However, the background would not have any physical singularity if the RG flow does not hit a 
singularity. 

One subtle point to identify the IR theory with the little string theory is that, in doing so, the 
Liouville direction must be a {\it space} direction, although at the beginning of the decay it is 
regarded as time direction. 
In fact, since there is a non-trivial linear dilaton background in the IR, the Wick rotation of the 
Liouville field would change the total central charge, that is, change the theory itself. 
Therefore, time direction has to be chosen from the other directions. 
If this should be taken seriously, it seems that, during the decay of the background, the time 
direction and a space direction should be 
exchanged, like a phenomenon which happens when an observer passes the event horizon of a black hole.

\appendix

\vspace{2cm}

\begin{flushleft}
{\bf \LARGE Appendix}
\end{flushleft}

\vspace{5mm}

\section{$T^2/{\bf Z}_3$ and LG} \label{LGspectrum}

\vspace{5mm}

In this appendix, it is shown that Type II string 
theory on $T^2/{\bf Z}_3$ and the LG 
model (\ref{LG}) have the same contents of tachyonic and massless states in the NS sector. 

To obtain the zero-point energy of each twisted sector in the orbifold theory, 
it would be the easiest way to employ the 
light-cone 
Green-Schwartz formalism, since in the NSR formalism 
the GSO-projection could be modified in the twisted sectors. 
For ${\bf C}/{\bf Z}_N$ orbifold with odd $N$, 
in which ${\bf Z}_N$ is defined not to produce a bulk tachyon, 
the zero-point energy of $k$-th twisted sector is 
\begin{equation}
E_0 = \left\{ 
\begin{array}{cc}
\displaystyle{\frac{k-N}{2N}}, & (k : \mbox{odd}), \\ [3mm] 
\displaystyle{-\frac k{2N}}, & (k : \mbox{even}). 
\end{array}
\right.
\end{equation}
In the $T^2/{\bf Z}_3$ orbifold, there are three fixed points which are locally ${\bf C}/{\bf Z}_3$, 
and the 
zero-point energies of the $k$-th twisted sector is 
\begin{equation}
E_0 = -\frac13
\end{equation}
for both $k=1,2$ and for all three fixed points. 
In the NSR formalism, 
this implies that in both twisted sectors the GSO-projection keeps the ground state of the Fock space, 
contrary 
to the untwisted sector. 

It is shown above that there are six tachyons with $\alpha'M^2/4=-1/3$ coming from the twisted sectors, 
and it can also be shown that 
there are 
eight massless states, two of which comes from the untwisted sector and the other six of which 
comes from the twisted sectors. 
All other states are massive. 

\vspace{3mm}

Primary fields $\Phi_{l,q}$ 
in ${\cal N}=2$ minimal model with level $k$ are labeled by two integers $l,q$ satisfying 
$0\le l\le k,\ -l\le q\le l$, and their 
weights and $U(1)$ charges are 
\begin{equation}
h_{l,q} = \frac{l(l+2)-q^2}{4(k+2)}, \hspace{5mm} Q_{l,q} = \frac q{k+2}. 
\end{equation}
For $k=1$, there are three primary fields $\Phi_{0,0}={\bf 1}$, $\Phi_{1,0},\ \Phi_{1,\pm1}$. 
If the minimal model is specified by $W(X)=X^3$, the field $X$ $(X^\dag)$ corresponds to $\Phi_{1,+1}$ 
$(\Phi_{1.-1})$, respectively, and all the other fields are obtained from $X,\ X^\dag$ via the operator 
product. 

Consider the LG model (\ref{LG}) which is the direct product of three ${\cal N}=2$ minimal models with 
$k=1$ described above. 
This model contains the following states
\footnote{The OPE of $XX^\dag$ contains {\bf 1} and operators with weights $h\ge1$.}, 
\begin{equation}
\begin{array}{rcl}
h-\frac12=-\frac12  & : & {\bf 1}, \\
-\frac13  & : & X,\ Y,\ Z,\ X^\dag,\ Y^\dag,\ Z^\dag, \\
-\frac16  & : & XY,\ X^\dag Y,\ XY^\dag,\ X^\dag Y^\dag,\ \cdots, \\
0         & : & XYZ,\ X^\dag YZ,\ XY^\dag,\ XYZ^\dag,\ X^\dag Y^\dag Z,\ X^\dag YZ^\dag,\ 
                XY^\dag Z^\dag,\ X^\dag Y^\dag Z^\dag. 
\end{array}
\end{equation}
One can show that $h-1/2>0$ for the other states in the LG model, that is, they are massive states 
for the corresponding string theory. 

One can see that states which are odd under 
\begin{equation}
X\ \to\ -X, \hspace{5mm} Y\ \to\ -Y, \hspace{5mm} Z\ \to\ -Z
   \label{GSO}
\end{equation}
correspond in one-to-one to the states in the orbifold theory. 
Note that not only weights but also $U(1)$ charges, which is related to the fermion number in the 
orbifold theory, coincide. 
Since massless states in the untwisted sector in the orbifold theory are (anti-)chiral primary states, 
they correspond to $XYZ,\ X^\dag Y^\dag Z^\dag$. 
The transformation (\ref{GSO}) can be regarded as the action of $(-1)^F$ on the LG superfields, and 
then the modding out the even states corresponds to the GSO-projection.

\end{document}